# Theoretical and Numerical Investigation of Liquid-Gas Interface Location of Capillary Driven Flow During the Time Throughout Circular Microchannels

Arshya Bamshad
Department of Mechanical Engineering, Faculty of Engineering
University of Central Florida
Orlando, USA
arshya.bamshad@knights.ucf.edu

Alireza Nikfarjam
Department of MEMS & NEMS, Faculty of New Sciences and Technologies
University of Tehran
Tehran, Iran
a.nikfarjam@ut.ac.ir

Mohammad Hossein Sabour
Department of Aerospace Engineering, Faculty of New Sciences and Technologies
University of Tehran
Tehran, Iran
sabourmh@ut.ac.ir

Hassan Raji
Department of Mechatronics Engineering, Faculty of New Sciences and Technologies
University of Tehran
Tehran, Iran
hraji@ut.ac.ir

*Abstract*—The main aim of this study is to find the best, most rapid, and the most accurate numerical method to find the liquid-gas interface of capillary driven flow during the time in circular Microchannels by using COMSOL Multiphysics software. Capillary driven flow by eliminating micropumps or any physical pressure gradient generators can make the microfluidic devices cheaper and more usable. Hence, by using this two-phase flow, the final costs of lots of microfluidic devices and lab-on-a-chip can significantly be decreased and help them to be commercialized. The first step to employing the capillary flow in these devices is the simulation of this flow inside the microchannels. One of the most common and valid software for this work is COMSOL Multiphysics; this fact reveals the importance of this study. In this research study, simulation results obtained by using two possible numerical methods in this software, for capillary flows of water and ethanol in two different circular microchannels, verified and compared with four other methods, which verified experimentally before. Finally, the most accurate and time-saving numerical method of this software will be specified. This appropriate technique can contribute to simulate microfluidic and lab-on-a-chip devices, which are made of different mechanical and electrical parts, in COMSOL Multiphysics software by choosing the best method.

I. INTRODUCTION

Liquid behavior inside a microchannel is an important factor for designing step in lab-on-chip devices. The science to investigate this fluid flow in microchannels is known as microfluidic [1]. Microfluidics is a multidisciplinary field that designs and manufactures practical systems, which uses small volumes of fluids. At least one of the dimensions of these systems should be less than 1 mm and more than 1 μm [1]. This field of science has been emerged in the 1980s to make development in inkjet print head, lab on chips, and micro devices [1]. Due to surface tension, energy dissipation, and fluid resistance, the behavior of the fluid flow in a microchannel is completely different from fluid flow in a macro channel [2-5]. This field studies the behavior of fluids inside the microchannels and finds new applications for it [2-5]. These microchannels can be manufactured in different methods with a variety of materials such as silicon, glass, polymers, etc. [6]

Microchannels, because of its high surface and small volume of the fluid can provide a lots of advantages such as high rate of heat and mass transfer, and so on [6]. According to these advantages, microchannels can implement in heat exchangers, bio-MEMS devices, Lab-on-a-chip devices, point of care diagnostics, and transport path of biological structures like DNA, and so on [1]. Because of the good match of these channels with scale of biological structures; furthermore, multiple functions for chemical analysis in a small area which can provide an appropriate efficiency, the microchannels are very popular [6]. There are different techniques to generate fluid flow inside microfluidic devices, i.e. pressure gradient generators, capillary driven flow, etc. Capillary flow, which is popular technique can emerge spontaneously in some polymers such as PMMA and glass made microchannels for bio-fluids [7, 8]. The other techniques to create fluid flow need some equipment like pumps, seal equipment, etc., which not only may increase the size of the microchips, but also can increase the cost of the product in comparison with the capillary-based microdevices [9]. Therefore, capillary driven flow can provide a lot of merits for implementing in the lab-on-a-chip or any micro-applicable devices. This multiphase flow caused when adhesive intermolecular forces of unlike molecules are more than cohesive intermolecular forces of similar molecules [9, 10]. Hence, the capillary force is a function of the substrate material, liquid type, temperature, and pressure of the experiment.

The exact location of the flow`s meniscus inside the microchannel during the time is one of the most important data of the capillary driven flow that plays a key role in design and analysis of the lab-on-a-chip and microchips. Depending on the liquid and substrate`s material of the microchannel, there are wettable and nonwettable surfaces. The wettable surface is a surface which the liquid can wet it. It means the contact angle for a liquid droplet on the surface is less than 90 degrees. But, the contact angle of the liquid droplet on the nonwettable surfaces is more than 90 degrees. The capillary flow only occurs in the wettable micro channels for fluids [8, 11]. Glass is a popular hydrophilic substrate, and a lot of experimental data exists about the capillary driven flow. Thus, it could be an appropriate material for analysis of different methods.

To sum up, by using the capillary driven flow, the cost of lab-on-a-chip, microfluidic-based devices, and MEMS devices will be declined, and these devices will be more usable for the people. This fact demonstrates the importance of an accurate simulation of capillary driven flow in the microchannels. Since, a lab-on-a-chip is constituted from different parts such as waveguides, microchannels, etc., it is necessary to use Multiphysics software to design and simulate whole parts of the microchip together. Such simulations are being used to determine any probable defeats in primitive steps of design. One of the most usable software to simulate lab-on-a-chip and microchips is the COMSOL Multiphysics. It follows that it is really important to know the most rapid computational technique with the least error to simulate the capillary driven flow via this software for finalizing the design. In what follows some of the most important equations to predict capillary driven flow are firstly discussed and then two possible solutions via COMSOL Multiphysics would be delineated. Finally, the errors and simulation times of these two COMSOL-based methods are compared with each other. The findings of this study can be applied for both horizontal and vertical microchannels in that it is assumed that the whole system is running at low Reynolds number in the microchannel [12].

## II. METHODOLOGY

In this paper, precise location of the liquid meniscus of capillary driven flow in different circular microchannels, over the time under negligible gravity condition were investigated. These locations by six different methods in two different circular glass microchannels were found within computer simulations. All of these techniques consider the capillary driven flow as a laminar flow because of its own low Reynolds number. This assumption is caused because of the negligible radius of the microchannels [13]; furthermore, this assumption is dictating that gravitational forces are negligible [12].

$$Re = \frac{\rho V D}{\mu} \qquad (1)$$

Where $\rho$ is the density of the fluid, V is the velocity of flow, D is the diameter of the microchannel, and $\mu$ is the dynamic viscosity of the fluid. As it was mentioned above, there are six different methods to find liquid-gas interface location in this study.

### A. Lucas-Washburn Equation

One of the most important solutions of capillary driven flow was solved by E. W. Washburn which is known as the Washburn equation. In addition, this equation was verified by experiment; consequently, is almost always taken for granted for explaining capillary driven flow. This equation derived from the Hagen-Poiseuille equation. According to the Hagen-Poiseuille equation, the volumetric flow rate (Q) in a circular channel has been defined as [4]:

$$Q = \frac{\rho \Delta P r^4}{8 \mu L} \qquad (2)$$

Where $\Delta P$ is the pressure gradient, L is the length of the liquid penetration into the microchannel, and r is radius of the microchannel. By using this equation, Washburn introduced an equation to predict the meniscus location during the time in a microchannel; based on this equation the meniscus`s location of a capillary flow at a circular microchannels defined as [14, 15]:

$$L^2 = \left(\frac{\sigma \cos\theta}{\mu \ 2}\right) r t \qquad (3)$$

Where $\sigma$ is surface tension, $\theta$ is contact angle, and t is time.

### B. Szekeley et al. Capillary Equation

This method is based on energy balance between kinetic and potential energies of the fluid in a control volume. It is a newer solution for capillary driven flow with less error in comparison with the Lucas-Washburn equation. This is a differential equation that is related to h, height of fluids in a vertical microchannel, and g is gravity; furthermore, the equation has been verified experimentally. According to this equation for a vertical microchannel [16]:

$$\left(h + \frac{7}{6} r\right)\frac{d^2 h}{dt^2} + 1.225 \left(\frac{dh}{dt}\right)^2 + C_1 h \frac{dh}{dt} = \frac{1}{\rho}[\Delta P - \rho g h] \quad (4)$$

$$C_1 = \frac{8\mu}{\rho r^2} \qquad (5)$$

$$\Delta P = \frac{2\sigma \cos\theta}{r} \qquad (6)$$

By disregarding the effect of gravity at low Reynolds flows in microchannels, the differential equation can be written for all circular microchannels as:

$$\sigma \pi D \cos\theta = \frac{\rho \pi D^2 q}{8}\left(\frac{dl}{dt}\right)^2 + \frac{\rho \pi D^2 (l+p)}{4}\frac{d^2 l}{dt^2} + 8\mu\pi l \frac{dl}{dt} \quad (7)$$

Where $q = 2.45$, and $p = \frac{7D}{12}$ for a fully developed flow.

### C. Batten Capillary Equation

Batten modified the equation (4) by doing some new experiments. He proposed new value 3.41 for the q parameter. By assuming this new value instead of 2.45, equation (7) transformed to the equation (8) [17].

$$\sigma \pi D \cos\theta = \frac{1.705 \rho \pi D^2}{4}\left(\frac{dl}{dt}\right)^2 + \frac{\rho \pi D^2 (l+\frac{7D}{12})}{4}\frac{d^2 l}{dt^2} + 8\mu\pi l \frac{dl}{dt} \quad (8)$$

### D. Ichikawa and Satoda Capillary Equation

With the advent of new detection technologies, Ichikawa et al. designed some experiments by using a high-speed camera to find the exact location of capillary flow`s meniscus in some PDMS and glass based microchannels for different liquids. Finally, they modified equation (4) by disregarding parameter p in the equation (4). In addition, they proposed new value for another parameter $q = 2$. Therefore, under these assumptions, equation (4) transformed to [18, 19]:

$$\left(\sigma \cos\theta + \frac{\rho g z\, D}{4}\right)\frac{\rho D}{\mu^2} - 8\frac{\rho D^2}{\mu}\frac{1}{D^2} l \frac{dl}{dt} =$$

$$\frac{1}{4}\left(\frac{\rho D^2}{\mu}\right)^2 \frac{1}{D^2}\frac{d}{dt}\left(s \frac{dl}{dt}\right) \qquad (9)$$

The equation (9) is known as one of the most accurate solutions for capillary driven flow in circular microchannels. [16] Finally, this equation under negligible gravity condition is transformed to:

$$\sigma \pi D \cos\theta = \frac{\rho \pi D^2}{4}\left(\frac{dl}{dt}\right)^2 + \frac{\rho l D^2}{4}\frac{d^2 l}{dt^2} + 8\mu l \frac{dl}{dt} \qquad (10)$$

### E. Level Set Method

In this method, the capillary driven flow in microchannels is considered as a laminar two phase flow. To find the accurate location of the meniscus in this method, the Navier-Stokes, Continuity, and the Level Set equations were solved together.

The fluid interface, the Level Set equation, of this two phase flow is defined as [20, 21]:

$$\frac{\partial \varphi}{\partial t} + u.\nabla\varphi = \gamma\nabla.(\varepsilon\nabla\varphi - \varphi(1-\varphi)\frac{\nabla\varphi}{|\nabla\varphi|}) \quad (11)$$

Where ε is interface thickness, and γ is re-initialization parameter that is 1 meter per second. In addition, φ is the level Set function, which it is 0 for the air and it is 1 for the water, and it is considered 0.5 for the liquid-air interface. The viscosity and the density are defined as [21]:

$$\rho = \rho_{air} + (\rho_{liquid} - \rho_{air})\varphi \quad (12)$$
$$\mu = \mu_{air} + (\mu_{liquid} - \mu_{air})\varphi \quad (13)$$

Also, delta function and normal of interface are defined as [22-24]:

$$\delta = 6|\varphi(1-\varphi)||\nabla\varphi| \quad (14)$$
$$n = \frac{\nabla\varphi}{|\nabla\varphi|} \quad (15)$$

The Navier-Stokes and the continuity equations are defined as [25-27]:

$$\rho\frac{\partial u}{\partial t} + \rho(u.\nabla)u = \nabla.[-pI + \mu(\nabla u + (\nabla u)^T)] + F_{st} + \rho g \quad (16)$$

$$\nabla.u = 0 \quad (17)$$

Where u denotes as velocity, p is pressure, I is the identity matrix, and $F_{st}$ represents the surface tension force acting at liquid-air interface. This parameter defines as [21, 28]:

$$F_{st} = \nabla.T \quad (18)$$
$$T = \sigma(I - (nn^T))\delta \quad (19)$$

There are some boundary conditions which are specific about our model. These boundary conditions are shown in figure 1.

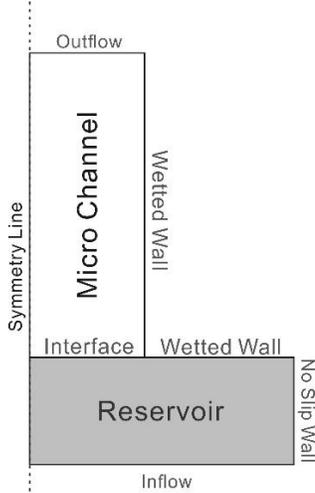

*Fig. 1 – The sketch of axisymmetric model of micro channel with its reservoir and its boundary conditions in the COMSOL Multiphysics*

The boundary conditions of wetted walls are defined as [28]:

$$u.n_{wall} = 0 \quad (20)$$
$$F_{fr} = -\frac{\mu}{\beta}u \quad (21)$$

Where $F_{fr}$ demonstrates frictional boundary force, and β indicates the slip length, which is equal to the mesh element size. [29, 30] Finally, the equations (11-21) were solved together.

*F. Phase Field Method*

Same as the previous sections, this method considers low Reynolds number for the capillary driven flow. To find the precise location of the meniscus in this method, Navier-Stokes, continuity, and the Cahn-Hilliard equations were solved together. The Cahn-Hilliard equation is defined as [31]:

$$\frac{\partial \varphi}{\partial t} + u.\nabla\varphi = \nabla.\frac{\gamma\lambda}{\varepsilon^2}\nabla\psi \quad (22)$$
$$\psi = -\nabla.\varepsilon^2\nabla\varphi + (\varphi^2 - 1)\varphi \quad (23)$$

Where ε is the interface thickness, λ is mixing energy density, γ is mobility, u is the fluid velocity, and φ is a dimensionless phase field variable in the liquid-gas interface, which is a number between -1 and 1. [32] Additionally, the surface tension coefficient is expressed as [27, 33]:

$$\sigma = \frac{2\sqrt{2}}{3}\frac{\lambda}{\varepsilon} \quad (24)$$

Volume fractions in the interface are defined as [21]:

$$V_{liquid} = \frac{1-\varphi}{2} \quad (25)$$
$$V_{air} = \frac{1+\varphi}{2} \quad (26)$$

Also, the total density and viscosity of the mixture are defined as [21, 33]:

$$\rho = \rho_{liquid} + (\rho_{air} - \rho_{liquid})V_{air} \quad (27)$$
$$\mu = \mu_{liquid} + (\mu_{air} - \mu_{liquid})V_{air} \quad (28)$$

In addition, surface tension force in the equation (16) is defined as [21]:

$$F_{st} = G\nabla\varphi \quad (29)$$
$$G = \lambda\left[-\nabla^2\varphi + \frac{\varphi(\varphi^2-1)}{\varepsilon^2}\right] = \frac{\lambda}{\varepsilon^2}\psi \quad (30)$$

Finally, in this method, the equations (16), (17), and (20-30) were solved together.

### III. RESULTS AND DISCUSSION

To numerically solve the equations of Level Set and Phase Field methods throughout the microchannels, 2D axisymmetric space in microfluidics module of COMSOL Multiphysics 4.4 – Finite Element Software – was used. In addition, transient study with phase initialization were selected as this module`s solver to simulate the capillary driven flow. Other information such as boundary conditions, mesh size, and mesh type are indicated in tables 1, and 2.

*Table 1 – Boundary conditions and values*

| The Boundary | Value |
|---|---|
| Inlet condition | $P = 1\ atm$ |
| Outlet condition | $P = 1\ atm$ |
| Interface thickness (Level Set Method) | $5\ \mu m$ |
| Interface thickness (Phase Field Method) | $6.5\ \mu m$ |

*Table 2 – Mesh type and size*

| Parameter | Value |
|---|---|
| Mesh type | Mapped |
| Maximum element size (*mm*) | 0.0065 |
| Minimum element size (*μm*) | 0.013 |
| Maximum element growth rate | 1.1 |
| Curvature factor | 0.2 |
| Resolution of narrow regions | 1 |

To investigate the accuracy of the simulation used via COMSOL Multiphysics (section 2.5 and 2.6) for the capillary driven flow in the circular microchannels, the meniscus`s location of the flow inside the microchannels were compared with the results of the meniscus`s location derived from other sections. Furthermore, to find the most rapid simulation method via COMSOL software, the simulation time of these two methods – section 2.5 and 2.6 – were compared with each other. To ensure the integrity of the obtained results, and also to increase the repeatability, the capillary driven flow of two different liquids – water and ethanol – in two circular microchannels with different diameters – 100 *μm* and 200 *μm* – were investigated. In addition, glass is considered as substrates of the microchannels. The physical characteristics of the liquids used in this study are provided in table 3.

*Table 3 – Physical Characteristics of the liquids*

| Liquid | Water | Ethanol |
|---|---|---|
| Density ($kg/m^3$) | 999.97 | 789.00 |
| Viscosity ($kg/m.s$) | 0.0010 | 0.0011 |
| Surface tension ($N/m$) | 0.072 | 0.022 |
| Contact angle (*degree*) | 40.0 | 67.5 |

Equations (7), (8), and (10) were solved numerically by using ODE 45 – MATLAB SIMULINK – and compared with other methods for water and ethanol for two different diameters of circular microchannels. The results are given in figures 2 through 13. A SUPERMICRO server computer with Intel Xenon CPU E5-2620 processor and 32 GB memory were implemented in this study. In the initial condition of the model, the reservoir is filled with liquid and the microchannel is filled with air, and then the microchannel begins to fill up with the liquid over time. Finally, all of these data were plotted for 0.1 seconds under negligible gravity condition.

Figures 2, 3, and 4 show the results of a capillary driven flow of water in a microchannel with 100 *μm* diameter.

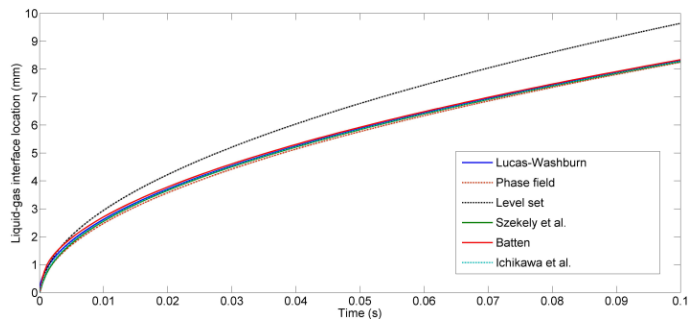

*Fig. 2 – Plot of interface location for water, D = 100 μm*

It is seen that the Level Set method is a proper method only for 0.004 seconds, and its error is growing during the time. Also, the Level Set method is diverging during the time, and the Phase Field method is converging to other methods.

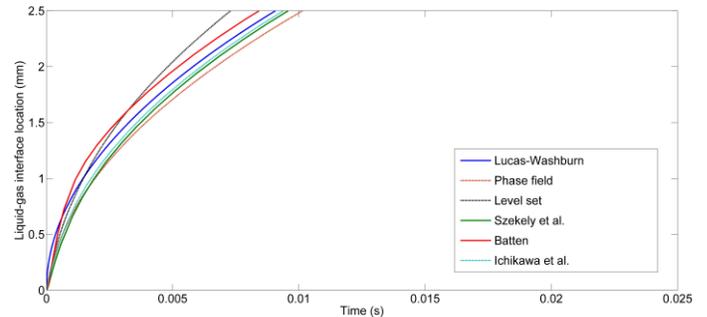

*Fig. 3 – Plot of interface location for water, D = 100 μm*

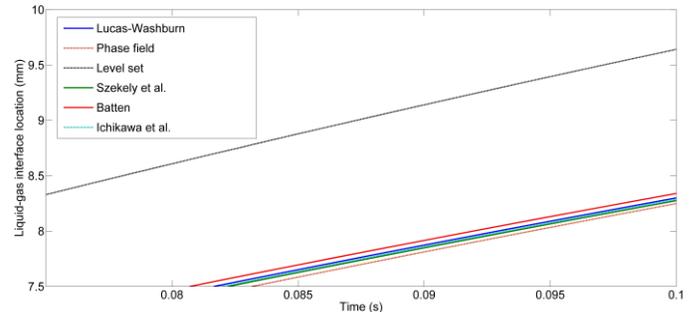

*Fig. 4 – Plot of interface location for water, D = 100 μm*

Figures 5, 6, and 7 demonstrate the result of the capillary flow of the water inside a microchannel with 200 *μm* diameter. It is observed that the Level Set method is only an appropriate technique for the first 0.002 seconds of the flow, and its own error is grew during the time.

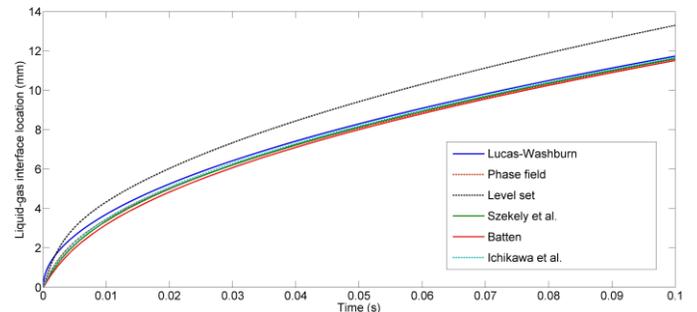

*Fig. 5 – Plot of interface location for water, D = 200 μm*

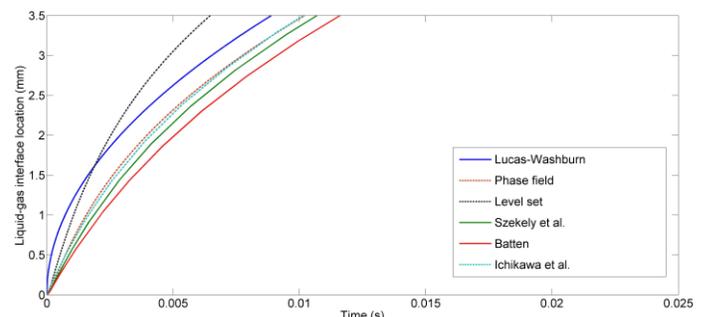

*Fig. 6 – Plot of interface location for water, D = 200 μm*

The Phase Field method is converging toward the other methods; however, the Level Set method is diverging.

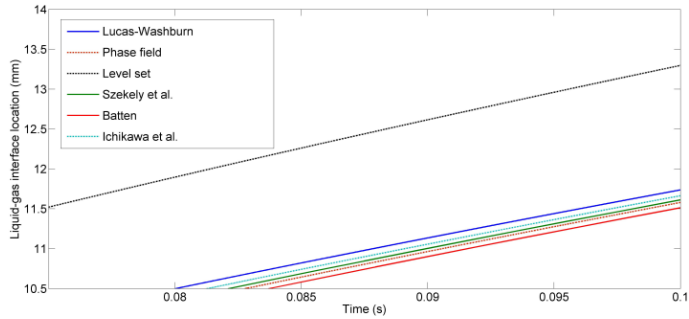

*Fig. 7 – Plot of interface location for water, D = 200 μm*

Figures 8, 9, and 10 indicate the results of the ethanol capillary flow in a microchannel with 100 μm diameter. It is discovered that the Level Set method is an apt technique merely for the first 0.005 seconds of the flow, and its own error is increasing during the time. But for more than this time, the Phase Field method is the more appropriate method in that it has converged toward the other methods over time.

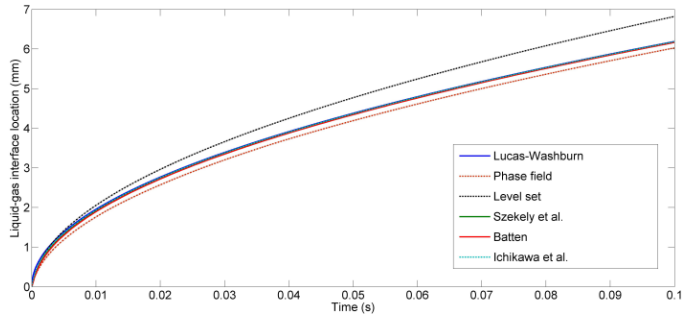

*Fig. 8 – Plot of interface location for ethanol, D = 100 μm*

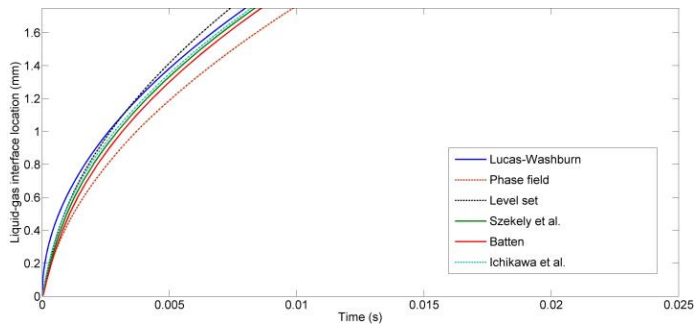

*Fig. 9 – Plot of interface location for ethanol, D = 100 μm*

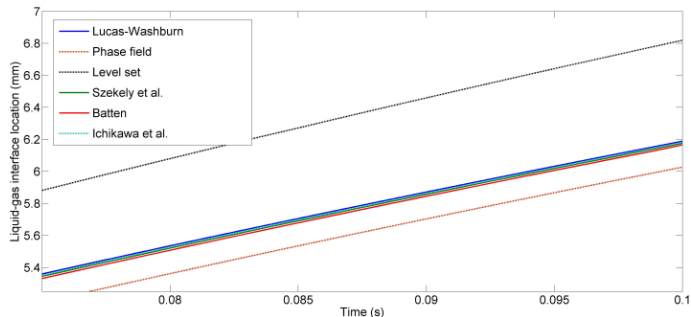

*Fig. 10 – Plot of interface location for ethanol, D = 100 μm*

Figures 11, 12, and 13 illustrate the results of ethanol capillary flow in a microchannel with 200 μm diameter. It is seen that the Level Set technique is a better method to simulate this flow for the first 0.002 seconds of the flow, but after this time, its error is grew over time. Consequently, the Phase Field method is the suitable method for a long period of time, since it is converging toward the other aforementioned methods.

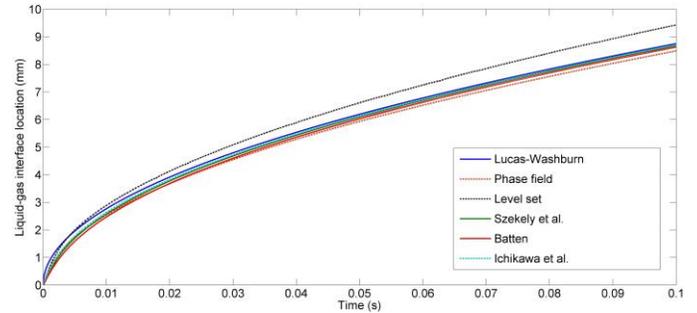

*Fig. 11 – Plot of interface location for ethanol, D = 200 μm*

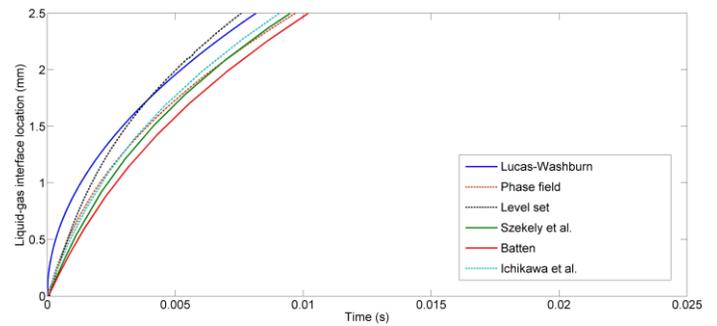

*Fig. 12 – Plot of interface location for ethanol, D = 200 μm*

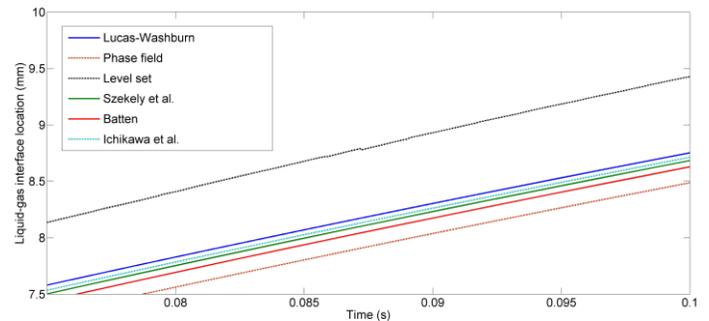

*Fig. 13 – Plot of interface location for ethanol, D = 200 μm*

As it was mentioned before, solution time and error of the capillary driven flows were investigated for the Level Set and Phase Field methods. Tables 4 and 5 contain errors of these two simulation methods in comparison with other aforementioned techniques for 0.1 seconds. These data demonstrate that the error of the Level Set method is much more than the Phase Field`s error.

The error of the liquid-gas interface`s location of the capillary driven flow of water, simulated by the Level Set method, is about 16 percent on average for a circular microchannel with 100 μm diameter. This amount is much more than the error of the Phase Field method, which is about 1 percent, for the same flow and microchannel. For a microchannel with 200 μm diameter, the error of the

meniscus`s location simulated by the Phase Field method is about 1 percent, which is much less than the error of the Level Set method that is about 14 percent of the water capillary driven flow.

Additionally, for the same microchannel with the Ethanol capillary driven flow, the error of the location of the liquid-gas interface simulated by the Level Set method is about 10 percent, which is much more than the error of the Phase Field method that is about 2 percent. The error of the Phase Field method for the Ethanol flow in the same microchannel is about 2 percent; this is much less than the error of the Level Set method, which is about 8 percent.

Also, solution time of the simulations for the Level Set and the Phase Field methods are presented in tables 6 and 7. According to these data, the computing time of the Phase Field method is almost half of the computing time of the Level Set method for the same simulation. Consequently, the Phase Field method is a more rapid technique than the Level Set method to simulate capillary driven flow in the circular microchannels.

*Table 4 – Errors of the Level Set and the Phase Field methods in comparison with other methods for water*

| Method | Diameter (μm) | Error (percent) | | | |
|---|---|---|---|---|---|
| | | Lucas-Washburn | Szekeley et al. | Batten | Ichikawa et al. |
| Level Set | 100 | 16.168 | 16.507 | 15.603 | 16.350 |
| | 200 | 13.287 | 14.510 | 15.492 | 14.002 |
| Phase Field | 100 | −0.612 | −0.322 | −1.095 | −0.456 |
| | 200 | −1.349 | −0.285 | 0.570 | −0.727 |

*Table 5 – Errors of the Level Set and the Phase Field methods in comparison with other methods for ethanol*

| Method | Diameter (μm) | Error (percent) | | | |
|---|---|---|---|---|---|
| | | Lucas-Washburn | Szekeley et al. | Batten | Ichikawa et al. |
| Level Set | 100 | 10.188 | 10.423 | 10.602 | 10.312 |
| | 200 | 7.713 | 8.562 | 9.257 | 8.199 |
| Phase Field | 100 | −2.622 | −2.415 | −2.256 | −2.512 |
| | 200 | −3.031 | −2.267 | −1.641 | −2.593 |

*Table 6 – Computing time by COMSOL Multiphysics 4.4 for the water*

| Method | Diameter (μm) | Time (seconds) |
|---|---|---|
| Level Set | 100 | 129736 |
| | 200 | 190643 |
| Phase Field | 100 | 66221 |
| | 200 | 107291 |

*Table 7 – Computing time by COMSOL Multiphysics 4.4 for the ethanol*

| Method | Diameter (μm) | Time (seconds) |
|---|---|---|
| Level Set | 100 | 79085 |
| | 200 | 91434 |
| Phase Field | 100 | 40883 |
| | 200 | 58862 |

## IV. CONCLUSION

To sum up, the Level Set method is a more accurate method to simulate capillary driven flow than the Phase Field method only for a short period of time, about 0.004 seconds, in the COMSOL Multiphysics software. But the Phase Field method is a more accurate and faster method than another above-mentioned method to find the meniscus`s location of the capillary driven flow for a long period of time. In fact, the meniscus`s location of capillary driven flows, simulated by the Level Set method, is diverging over time, but the derived data of the Phase Field method are converging throughout time toward other previously verified solutions. Furthermore, the computing time of the Level Set method is two times greater than the computing time of the Phase Field method. To find the exact location of the liquid-gas interface during the time of this two-phase flow, the Phase Field method is a better technique than the Level Set method.

Hence, to simulate capillary-based lab-on-a-chip and microdevices by COMSOL Multiphysics, which is the most popular software for simulating lab-on-a-chip, the Phase Field method ought to be implemented. This fact contributes to improving design and simulation of capillary-based microdevices and lab-on-a-chip because of its aforementioned advantages – its decrease in error and computing time.

ACKNOWLEDGMENT

The authors are grateful for the help of Mrs. Hajar Mahvi during this work.